\documentclass[letterpaper, 10pt, conference]{ieeeconf}

\usepackage[british]{babel}
\usepackage{amsmath,amssymb,amsfonts, bm}
\usepackage{graphicx}
\usepackage{textcomp}
\usepackage{float}
\usepackage{comment}

\usepackage{mathrsfs}
\usepackage[dvipsnames]{xcolor}
\usepackage[acronym]{glossaries}
\usepackage{lipsum}
\usepackage{algorithm}
\usepackage{algpseudocode}

\newcommand{\Pdist}{\mathscr{P}}
\newcommand{\Borel}{\mathcal{B}}
\renewcommand{\Pr}{\mathbb{P}}
\newcommand{\rel}{\mathcal{R}}

\newcommand{\C}{\mathbf{C}}
\newcommand{\M}{\mathbf{M}}
\newcommand{\X}{\mathbb{X}}
\newcommand{\T}{\mathbb{T}}
\newcommand{\U}{\mathbb{U}}
\newcommand{\Y}{\mathbb{Y}}

\newcommand{\GaussMix}{\mathcal{N}\hspace{-0.79em}\mathcal{N}}

\makeatletter
\newcommand*\bigcdot{\mathpalette\bigcdot@{.5}}
\newcommand*\bigcdot@[2]{\mathbin{\vcenter{\hbox{\scalebox{#2}{$\m@th#1\bullet$}}}}}
\makeatother

\newtheorem{theorem}{Theorem}
\newtheorem{remark}{Remark}
\newtheorem{definition}[theorem]{\noindent Definition}

\newtheorem{proposition}[theorem]{Proposition}
\newtheorem{corollary}[theorem]{Corollary}

\allowdisplaybreaks

\IEEEoverridecommandlockouts   

\usepackage{xspace}
\usepackage{cite}
\usepackage{array}
\usepackage{color}
\usepackage{balance}
\usepackage{booktabs}
\usepackage{mathtools}
\usepackage{wrapfig}
\usepackage{upgreek}
\usepackage{capt-of}

\begin{document}

\title{Stochastic MPC for Finite Gaussian Mixture Disturbances with Guarantees
\thanks{This work is supported by the Dutch NWO Veni project CODEC under grant number 18244 and the European project SymAware under grant number 101070802. $^1$Department of Electrical Engineering (Control Systems Group), Eindhoven University of Technology, The Netherlands. Emails:\{m.h.w.engelaar, m.lazar, s.haesaert\}@tue.nl \& swaanen@kth.se.}
}

\author{M. H. W. Engelaar$^1$, M. P. P. Swaanen$^1$, M. Lazar$^1$, and S. Haesaert$^1$}

\maketitle

\begin{abstract}

This paper presents a stochastic model predictive control (SMPC) algorithm for linear systems subject to additive Gaussian mixture disturbances, with the goal of satisfying chance constraints. We focus on a special case where each Gaussian mixture component has a similar variance. To solve the SMPC problem, we formulate a branch model predictive control (BMPC) problem on simplified dynamics and leverage stochastic simulation relations (SSR). Our contribution is an extension of the SMPC literature to accommodate Gaussian mixture disturbances while retaining recursive feasibility and closed-loop guarantees. We illustrate the retention of guarantees with a case study of vehicle control on an ill-maintained road.

\end{abstract}


\section{Introduction}

Control theory, a fundamental discipline in engineering and applied mathematics, offers a broad spectrum of techniques, ranging from simple, intuitive methods to highly sophisticated and computationally intensive approaches \cite{Guo2022}. Within this spectrum, stochastic model predictive control (SMPC) is a robust method for effectively managing chance constraints—set constraints that must be met with a specified probability—while addressing uncertainties within dynamical systems, often described by probability distributions \cite{Mesbah2016}. SMPC has been effectively utilized in various applications, including vehicle path planning, air traffic control, building climate control, and operations research and finance \cite{Mesbah2016}.

SMPC approaches are typically classified into two broad categories: randomized and analytic approximation methods \cite{Farina2016}. Randomized methods rely on generating realizations of disturbances, while analytic approximation methods generally seek to convert the stochastic control problem into a deterministic one. The analytic approximation category has, amongst others, work dedicated to uncertainty with bounded support \cite{Kouvaritakis2016}, identifying control-invariant sets \cite{kuwata2009real}, and analytic and data-driven scenarios \cite{ren2024recursively}. Among the literature, many consider guarantees on recursive feasibility and satisfaction of chance constraints of primary importance \cite{Kouvaritakis2016}.

For a subclass of stochastic control problems with unbounded uncertainty—modelled by central convex unimodal distributions—researchers have developed SMPC algorithms using analytic approximation \cite{Hewing2018,Mark2019distributed,Schluter2022,Engelaar2023stochastic} to ensure recursive feasibility and chance constraint satisfaction. However, these approaches struggle with more complex distributions. Central convex unimodality is crucial for enforcing chance constraints in closed-loop dynamics but often fails in real-world scenarios. A more practical model involves Gaussian mixtures, which can approximate any continuous distribution with arbitrary precision \cite{Goodfellow-et-al-2016,Maz1996approximate}.

Approximating distributions in the context of predictive control have been considered in \cite{Coulson2021distributionally} in which distributional robust data-enabled predictive control was used. Herein, a ball is constructed in the space of probability distributions centred around the uniform distribution of the sample set, which contains the actual distribution with a predefined probability \cite{Mohajerin2018data}. The control strategy is developed under the assumption of the worst-case distribution scenario. Similarly, Gaussian mixtures in the context of predictive control have been considered in \cite{weissel2009stochastic}, which assumed non-linear dynamics subject to additive disturbances. Nevertheless, tractable implementation, deriving closed-loop guarantees, and establishing recursive feasibility remain open research problems. Recent work \cite{ren2024recursively}, however, did establish (restrictive) recursive feasibility and closed-loop guarantees for Gaussian mixtures.

\begin{figure}[htp]
	\centering
	\includegraphics[width=\columnwidth]{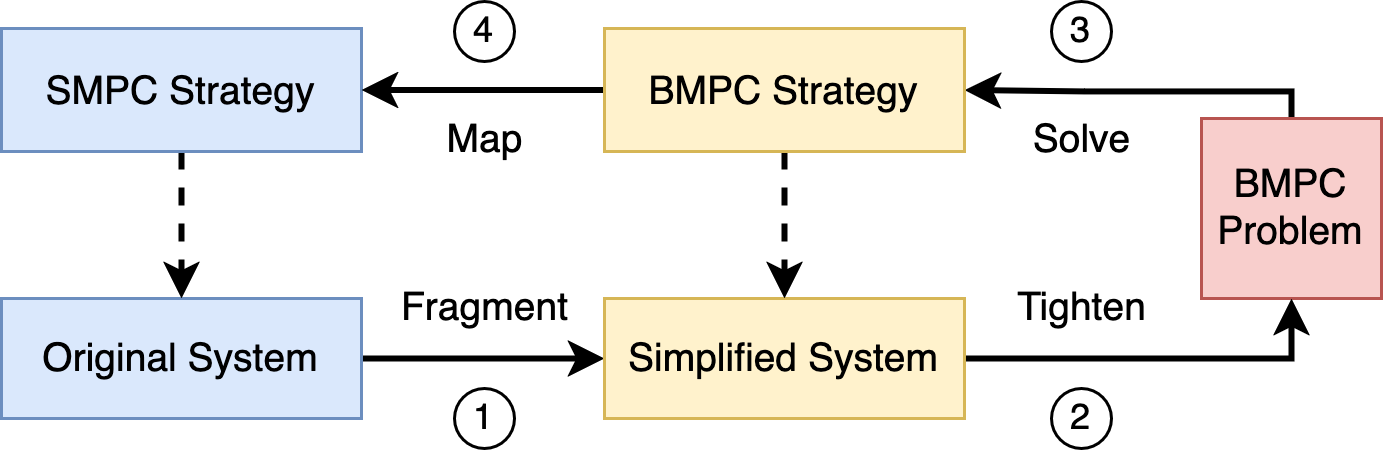}
	\vspace{-0.9em}
	\caption{A high-level illustration of the approach. (1) fragment the system, (2) tighten the chance constraints, (3) solve the BMPC problem, and (4) map the BMPC strategy to an SMPC strategy.}
	\label{Fig:Scheme}
\end{figure}

This paper develops an SMPC algorithm for linear systems subject to additive Gaussian mixture disturbances, aiming to establish recursive feasibility and satisfy chance constraints. We focus on a special case where each Gaussian mixture component has similar variance. The approach, depicted in Fig. \ref{Fig:Scheme}, begins by fragmenting the system dynamics through disturbance decoupling—separating the mixture into discrete and continuous components—and system decomposition—partitioning the dynamics into nominal and error parts. Next, a branch model predictive control (BMPC) problem on the simplified dynamics is derived by tightening the chance constraints. Finally, a control refinement operator (CR-operator), obtained through stochastic simulation relations (SSR), maps the BMPC strategy, the control strategy obtained from solving the BMPC problem, to an SMPC strategy, a control strategy that solves the SMPC problem.

The primary contribution of this paper is extending SMPC methods from \cite{Hewing2018, Engelaar2023stochastic} to handle additive Gaussian mixture disturbances while preserving key properties such as recursive feasibility and closed-loop guarantees. Our approach differs from existing Gaussian mixture SMPC literature, such as \cite{ren2024recursively}, by leveraging SSR to simplify system dynamics before control synthesis. A key limitation of our work is the assumption of a special case of the Gaussian mixture; however, we argue that this restriction can be mitigated at the cost of increased conservatism during control synthesis.



\section{Preliminaries \& Problem Statement} \label{Sec:Prel}

For a given probability measure $\mathbb P$ defined over Borel measurable space $(\X,\mathcal{B}(\X))$, we denote the probability of an event $\mathcal{A} \in \Borel(\X)$ as $\mathbb{P}(\mathcal{A})$. We denote the set of all probability measures over $(\X,\mathcal{B}(\X))$ by $\Pdist(\mathbb{X})$. In this paper, we will work with Polish spaces and Borel measurability. Further details on measurability are omitted, and we refer the interested reader to \cite{Bertsekas1996}. The vector of all elements one or zero is denoted, respectively, by $\boldsymbol 1_n, \boldsymbol 0_n \in \mathbb{R}^n$, and $\boldsymbol 1$ or $\boldsymbol 0$ if the context is clear. The Minkowski set difference and sum of $\mathcal{A},\mathcal{B} \subseteq \mathbb{R}^n$ are, respectively, given by $\mathcal{A} \ominus \mathcal{B} := \{a \mid a+b \in \mathcal{A}, \ \forall b \in \mathcal{B}\}$ and $\mathcal{A} \oplus \mathcal{B} = \{ a + b \mid \forall a \in \mathcal{A}, \ \forall b \in \mathcal{B} \}$.


\subsection{Stochastic System Dynamics}

We consider linear dynamics with additive noise, given by
\begin{equation} \label{Sys}
	x(k+1)=Ax(k)+Bu(k)+w(k),
\end{equation}
where $(A,B)$ is stabilizable, $x(k) \in \mathbb{R}^n$ is the state of the system, $x(0) \in \mathbb{R}^n$ is an initial state, $u(k) \in \mathbb{R}^m$ is the input of the system and $w(k) \in \mathbb{R}^n$ is an independent, identically distributed noise disturbance with distribution $\mathcal{Q}_w$, i.e.,  $w(k) \sim \mathcal{Q}_w$. We will assume that the distribution $\mathcal{Q}_w$ is a mixture of Gaussians denoted by $\GaussMix(\boldsymbol \mu, \boldsymbol \pi, \Sigma)$ with $\boldsymbol \mu = \{\mu_1, \cdots, \mu_L\}$, $\mu_i \in \mathbb{R}^n$, $\boldsymbol \pi = \{\pi_1, \cdots, \pi_L\}$, $\pi_i \in [0,1]$ and $\Sigma \in \mathbb{R}^{n \times n}$ a strictly positive definite matrix. The probability density function of $\GaussMix(\boldsymbol \mu, \boldsymbol \pi, \Sigma)$ is defined as
\begin{equation} \label{Mixture}
    \textstyle g(x;\boldsymbol \mu, \boldsymbol \pi, \Sigma):= \sum_{i=1}^L \pi_i g(x;\mu_i,\Sigma),
\end{equation}
where $\sum_{i=1}^L \pi_i=1$, and $g(x;\mu_i,\Sigma)$ is the probability density function of a multivariate Gaussian distribution. All multivariate Gaussian distribution components within the mixture have different means but share the same variance.

To control the system, we define a sequence of policies $$\boldsymbol f:= \{f_0, f_1,\dots\},$$ such that $f_k: \mathbb{H}_k \to \mathbb{R}^m$ maps history to inputs. The history is defined as $\mathbb{H}_k:=(\mathbb{R}^n \times \mathbb{R}^m)^k \times \mathbb{R}^n$, with elements $\eta(k):=(x(0),u(0), \cdots, u(k-1), x(k))$, representing the previous states and inputs. By implementing control strategy $\boldsymbol f$ upon system \eqref{Sys}, we obtain its controlled form for which the control input satisfies the feedback law $u(k)= f_k(\eta(k))$ with $\eta(k) \in \mathbb{H}_k$. We indicate the input sequence of system \eqref{Sys} by $\boldsymbol u:=\{u(0),u(1),\dots\}$ and define its executions as sequences of states $\boldsymbol x:=\{x(0),x(1),\dots\}$. We define the suffix of any sequence $\boldsymbol x$ by $\boldsymbol{x}_k=\{x(k), x(k+1), \dots\}$. Any sequence $\boldsymbol{x}$ can be interpreted as a realization of the probability distribution induced by implementing control strategy $\boldsymbol{f}$ upon the system \eqref{Sys}, denoted by $\boldsymbol{x} \sim \mathbb{P}_{\boldsymbol{f}}$.

\begin{remark}
	Unlike general Gaussian mixtures, where each component may have a different variance, our approach assumes all components share the same variance. While this may seem restrictive, any Gaussian mixture can be considered by over-approximating its components with a single Gaussian distribution, returning a mixture as in \eqref{Mixture}. However, because we are over-approximating the actual disturbance, any control strategy synthesized will be conservative.
\end{remark}


\subsection{Chance Constraints}

In this paper, the objective is to synthesize a control strategy $\boldsymbol f$ such that, if implemented, satisfaction is guaranteed for a given specification. This paper defines the specifications as set constraints on the state and input of system \eqref{Sys}, where satisfaction is to be guaranteed with predefined probability at each time instance. The following chance constraints describe such a specification.
\begin{subequations} \label{ChanCons}
	\begin{align}
		\mathbb{P}_{\boldsymbol f}(x(k) \in \mathcal{X} \mid x(0)) &\geq p_x, \ k \in \mathbb{N}-\{0\}, \label{ChanCons_x} \\
		\mathbb{P}_{\boldsymbol f}(u(k) \in \mathcal{U} \mid x(0)) &\geq p_u, \ k \in \mathbb{N}, \label{ChanCons_u}
	\end{align}
\end{subequations}
where $\mathcal{X}$ and $\mathcal{U}$ are convex sets, and $p_x$ and $p_u$ represent the \emph{target lower bounds}, i.e., the minimal predefined probability targets of the set constraints. It is assumed that the convex sets $\mathcal{X}$ and $\mathcal{U}$ contain the origin within their interior. The chance constraints are defined with respect to the initial state, i.e., conditioned based on the initial state $x(0)$.


\subsection{Problem Formulation}

\begin{figure}[htp]
    \centering
    \includegraphics[width=\columnwidth]{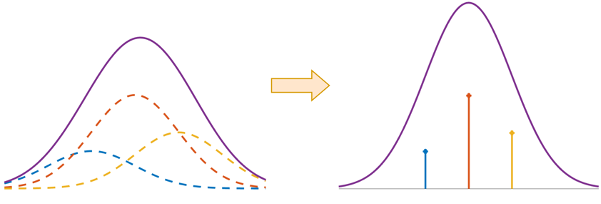}
    \vspace{-1.5em}
    \caption{An illustration of the decoupling of a Gaussian mixture (Left: Purple) into a continuous (Gaussian) distribution (Right: Purple) and a discrete distribution (Right: Stems). The dashed lines represent the Gaussian components of the mixture. Each stem corresponds to a component.}
    \label{Fig:Mixture}
\end{figure}

This paper aims to synthesize a control strategy $\boldsymbol f$, using stochastic model predictive control, such that stochastic linear system \eqref{Sys} satisfies the chance constraints \eqref{ChanCons}. As a secondary objective, we want to update the control strategy as new measurements become available. The corresponding SMPC problem can be formulated as
\begin{subequations} \label{SMPC}
	\begin{align}
		&\textstyle \min _{\boldsymbol f_k} \quad J(\boldsymbol x_k, \boldsymbol u_k), \quad \text{s.t.} \\
		&x(k+i+1) = Ax(k+i) +Bu(k+i) +w(k+i),\\ 
		& x(k+i+1), u(k+i) \text{ satisfy } \eqref{ChanCons}, \ \forall i \in \mathbb{N},
	\end{align}
\end{subequations}
where $J$ is some cost function. 
A key challenge arises from the Gaussian mixture disturbance, which provides a more realistic model than a standard Gaussian but significantly complicates the SMPC problem. To address this, we aim to decouple the mixture disturbance into discrete and continuous components (Fig. \ref{Fig:Mixture}) and utilize constraint tightening and control refinement to obtain a (suboptimal) SMPC strategy.

\section{Reformulating the SMPC Problem} \label{Sec:Main_Sys}

In synthesizing an SMPC strategy, the Gaussian mixture presents a key challenge. To address this, we simplify the system dynamics \eqref{Sys} by decoupling the Gaussian mixture and decomposing the dynamics. This results in a simplified model, referred to as the abstract system, whose dynamics align with existing works \cite{Hewing2018,Engelaar2023stochastic}, allowing us to build upon established SMPC methods. In this section, we first construct the abstract system and then leverage stochastic simulation relations to establish a control refinement operator that maps any control strategy designed for the abstract system to one for the original system \eqref{Sys}. We formally prove that this CR-operator preserves chance constraint satisfaction.


\subsection{Abstract System Design}

To design the abstract system, we first decouple the mixture into a discrete and continuous distribution, obtaining the dynamical system given by
\begin{equation} \label{SysNew}
	x(k+1) = Ax(k)+Bu(k)+w_x(k)+w_e(k),
\end{equation}
where $w_x(k) \sim \mathcal{Q}_w^x$ is a discrete distribution with probability mass function
\begin{equation} \label{Eq:DisDis}
	g(w_x ; \boldsymbol \mu, \boldsymbol \pi)=\begin{cases}
		\pi_i &\text{ if } w_x = \mu_i,\\
		0 &\text{ otherwise},
	\end{cases}
\end{equation}
and $w_e(k) \sim \mathcal{Q}_w^e:=\mathcal{N}(\boldsymbol 0, \Sigma)$. Next, we decompose dynamics \eqref{SysNew} into a nominal and an error part similar to \cite{Hewing2018}, i.e., $x(k)=z(k)+e(k)$. The nominal dynamics, denoted as $z$, contain stochasticity with finite support, and the error dynamics, denoted as $e$, are autonomous and contain stochasticity with infinite support. The abstract system is then given by the stochastic linear system
\begin{subequations} \label{AbsSys}
	\begin{align}
		z(k+1) & = Az(k)+Bv(k) + w_x(k), \label{Eq:Nom}\\
		e(k+1) &= A_Ke(k)+w_e(k), \label{Eq:Err}
	\end{align}
\end{subequations}
where $v(k)$ is the nominal input, $A_K=A+BK$ and $K$ is a stabilizing feedback gain meant to keep the error $e$ small. $K$ exists since $(A,B)$ is stabilizable. We note that the dynamical systems \eqref{SysNew} and \eqref{AbsSys}, for the same disturbance $w_x(k)$ and $w_e(k)$, satisfy $x(k)=z(k)+e(k)$ if
\begin{equation} \label{RelInput}
	v(k) = u(k) - Ke(k).
\end{equation}

\begin{remark}
	Systems \eqref{Sys} and \eqref{AbsSys} are not equivalent. Therefore, proving that a controller designed for system \eqref{AbsSys} translates to a valid controller for system \eqref{Sys} is non-trivial.
\end{remark}


\subsection{Stochastic Simulation Relation}

To design a CR-operator that preserves chance constraint satisfaction, we employ stochastic simulation relations. To define stochastic simulation relations, the notion of Markov decision process and lifting are required \cite{Haesaert2017verification}.

\begin{definition}[Markov Decision Process] \label{Def:Markov}
	The tuple $\M=(\X,\T,\U,h,\Y)$ defines a Markov decision process (MDP). A MDP is characterized by 1) Polish sets $\X$ and $\U$; by 2) $\T$, a Borel measurable stochastic transition kernel that assigns to each state $x \in \X$ and input $u \in \U$, a probability measure $\T(\cdot {\mid} x,u) \in \Pdist(\X)$; and by 3) $h:\mathbb{X} \to \mathbb{Y}$, a measurable output map that assigns to each state $x \in \mathbb{X}$ an output $y \in \mathbb{Y}$.
\end{definition}

The MDP representation is often non-unique. In this paper, we consider the following MDP representations, respectively, for original system \eqref{Sys} and abstract system \eqref{AbsSys}:
\begin{equation}
	\M := \begin{cases}
		\mathbb{X} := \mathbb{R}^n \times \mathbb{R}^m, \ \mathbb{Y} := \mathbb{R}^n \times \mathbb{R}^m, \ \mathbb{U} := \mathbb{R}^m, \\
		h(x,x_i) := (x,x_i), \ \mathbb{T} (dx \times dx_i \mid x,u) \\
		:= \sum_{i=1}^L \pi_i \mathcal{N}(dx;Ax+Bu+\mu_i,\Sigma) \delta(dx_i-u). \nonumber
	\end{cases}
\end{equation}
\begin{equation}
	\hat\M := \begin{cases}
		\hat{\mathbb{X}} := \mathbb{R}^n \times \mathbb{R}^n \times \mathbb{R}^m, \ \hat{\mathbb{Y}} := \mathbb{R}^n \times \mathbb{R}^m, \ \hat{\mathbb{U}} := \mathbb{R}^m, \\
		\hat{h}(z,e,\hat x_i) := (z+e,\hat x_i), \ \mathbb{T} (dz \times de \times d\hat x_i \mid z,e,v) \\
		:= \sum_{i=1}^L \pi_i \delta(dz-Az-Bv-\mu_i) \mathcal{N}(de;A_Ke,\Sigma) \\
		\delta(d\hat x_i-v-Ke). \nonumber
	\end{cases}
\end{equation}
We exclude the time indexes for brevity and note that each (sequence) element $s$ has index $k$, and each differential $ds$ has index $k+1$. $\delta$ denotes the Dirac delta distribution.

For MDP $\M$ and $\hat{\M}$, the transition kernels depict, respectively, the system dynamics of \eqref{Sys} and \eqref{AbsSys}, with an additional intermediate state element $x_i(k), \hat x_i(k)$ representing, respectively, the input $u(k-1)$ and the input relation $v(k-1)+Ke(k-1)$. Both of these are also our secondary outputs together with the primary outputs of the state $x(k)$ for $\M$ and the state relation $z(k)+e(k)$ for $\hat{\M}$. The sequel will show that the output pairs $x(k)$ and $z(k)+e(k)$, and $u(k)$ and $v(k)+Ke(k)$ have similar probability when the control inputs of both systems satisfy certain relations. 

\begin{definition}[Lifting] \label{Def:lifting}
	Consider two measurable spaces $(\mathbb{X},\mathcal{B}(\mathbb{X}))$ and $(\hat{\mathbb{X}},\mathcal{B}(\hat{\mathbb{X}}))$, and a relation $\rel\subseteq \mathbb{X}\times \hat{\mathbb{X}}$ with $\rel\in\mathcal B(\mathbb{X}\times \hat{\mathbb{X}})$. Pairs of probability measure $(\mathbb{P}_1, \mathbb{P}_2) \in \Pdist(\mathbb{X}) \times \Pdist(\hat{\mathbb{X}})$ belong to the lifted relation $\bar\rel$ if there exists a probability measure $\mathbb{W}$, referred to as a lifting, over a measurable space $(\mathbb{X}\times \hat{\mathbb{X}},\mathcal B(\mathbb{X}\times \hat{\mathbb{X}}))$ such that{ \setlength{\parskip}{1pt}\setlength{\parsep}{1pt}
		\begin{description}
			\item[\normalfont\textbf{I.}] $\mathbb W( \mathcal{A}_1\times \hat{\mathbb{X}})=\mathbb P_1( \mathcal{A}_1)$ for all $\mathcal{A}_1\in \mathcal{B}(\mathbb{X})$;
			\item[\normalfont\textbf{II.}] $\mathbb W( \mathbb{X}\times \mathcal{A}_2)=\mathbb P_2(\mathcal{A}_2)$ for all $\mathcal{A}_2\in \mathcal{B}(\hat{\mathbb{X}})$;
			\item[\normalfont\textbf{III.}] $\mathbb{W}\left(\rel\right)=1$.
	\end{description}}
\end{definition}

The interpretation of the lifting $\mathbb{W}$ is a relation between realizations of probability measures within the confine of a relation $\mathcal R$. Lifting also generalizes the notion of coupling as explained in \cite{Haesaert2017verification}. Utilizing the previous definitions, we obtain a definition for stochastic simulation relations \cite{Haesaert2017verification}.

\begin{definition}[Stochastic Simulation Relation]\label{Def:SSR}
	The MDP $\hat{\M}=(\hat\X,\hat\T,\hat\U,\hat h,\hat\Y)$ is simulated by MDP $\M=(\X,\T,\U,h,\Y)$ if there exists a Borel measurable interface function $\nu:\hat \U\times\hat \X\times\X \to \mathbb{U}$ and a relation $\rel\subseteq \hat \X\times \X$ with $\rel\in\mathcal B(\hat \X\times \X)$ for which there exists a Borel measurable stochastic kernel $\mathbb{W}(\,\cdot\,{\mid} \hat u,\hat x,x)$ on $\hat \X\times\X$ given values in $\hat \U\times\hat \X\times\X$, such that
	{\setlength{\parskip}{-1pt}\setlength{\parsep}{0pt}
		\begin{enumerate}
			\item $\forall (\hat x,x)\in \rel$, $ \hat h(\hat x)=h(x)$;
			\item $\forall (\hat x,x)\in \rel$, $\forall {\hat u}\in \hat \U$, $(\hat{\mathbb T}(\cdot| \hat x,\hat u),\mathbb T(\cdot| x,\nu(\hat u,\hat x,x)) \in \bar \rel$ with lifted probability measure $\mathbb{W}(\cdot \mid \hat u,\hat x,x)$.
	\end{enumerate}}
\end{definition}

The interpretation of the above conditions is that 1) state components in the relation behave similarly and 2) that for any element in the relation and any input on the simulated system, any realization of subsequent states, connected via the lifting and interface function, will be contained in the relation, making the relation invariant over the transition kernels. Utilizing the above definition, it can be proven that the original system $\M$ simulates the abstract system $\hat{\M}$.
\begin{theorem} \label{Thm:SSR}
	System $\hat{\M}$ is simulated by $\M$ via relation $\mathcal{R}=\{(z,e,\hat x_i,x,x_i) \mid x=z+e, x_i=\hat x_i\}$ and interface function $\nu(e,v)=v+Ke$.
\end{theorem}

\begin{proof}
	First, we consider a relation between the disturbances $w$, $w_x$ and $w_e$, given by $\mathcal{R}_1:=\{(w,w_x,w_e) \mid w=w_x+w_e\}$. We claim that distributions $\mathcal{Q}_w$ and $\mathcal{Q}_w^x \otimes \mathcal{Q}_w^e$ belong to the lifted relation. To prove this claim, we consider the potential lifting given by
	\begin{align} \label{Eq:Lift}
		\textstyle \mathbb{W}_1(w,w_x,w_e; \boldsymbol \mu, \boldsymbol \pi, \Sigma) &= \GaussMix(w;\boldsymbol \mu, \boldsymbol \pi, \Sigma) \bigcdot \nonumber\\
		\textstyle\sum_i\left[ \delta(w_x-\mu_i)\frac{\pi_i \mathcal{N}(w; \mu_i, \Sigma)}{\GaussMix(w;\boldsymbol \mu, \boldsymbol \pi, \Sigma)}\right] &\bigcdot \delta(w_e-w+w_x),
	\end{align}
	where $\delta$ denotes the Dirac delta distribution. To ascertain the validity of the lifting, we consider Def. \ref{Def:lifting}. First, we note that $\mathbb{W}_1\left(\rel_1\right)=1$ due to the second Dirac delta. Regarding the requirements I and II, we consider two integrals given by
	\begin{align*}
		\textstyle\iint_{w_x,w_e} &\mathbb{W}_1(w,w_x,w_e; \boldsymbol \mu, \boldsymbol \pi, \Sigma)dw_x dw_e, \\
		\textstyle\int_{w} &\mathbb{W}_1(w,w_x,w_e; \boldsymbol \mu, \boldsymbol \pi, \Sigma)dw.
	\end{align*}
	Solving the former, the result is given by 
	\begin{align*}
		&\textstyle\iint_{w_x,w_e} \mathbb{W}_1(w,w_x,w_e; \boldsymbol \mu, \boldsymbol \pi, \Sigma)dw_x dw_e = \GaussMix(w;\boldsymbol \mu, \boldsymbol \pi, \Sigma),
	\end{align*}
	while the latter will result in
	\begin{align*}
		&\textstyle\int_{w} \mathbb{W}_1(w,w_x,w_e; \boldsymbol \mu, \boldsymbol \pi, \Sigma)dw = \int_{w} \GaussMix(w;\boldsymbol \mu, \boldsymbol \pi, \Sigma) \bigcdot\\
		&\textstyle\sum_i\left[ \delta(w_x-\mu_i)\frac{\pi_i \mathcal{N}(w; \mu_i, \Sigma)}{\GaussMix(w;\boldsymbol \mu, \boldsymbol \pi, \Sigma)}\right] \bigcdot \delta(w_e-w+w_x) dw =\\
		&\textstyle\int_w \sum_i\left[ \delta(w_x-\mu_i) \pi_i \mathcal{N}(w; \mu_i, \Sigma)\right] \delta(w_e-w+w_x) dw =\\
		&\textstyle\sum_i \pi_i \delta(w_x-\mu_i) \mathcal{N}(w_e; \mu_i-w_x, \Sigma) = \\
		&\textstyle\sum_i \pi_i \delta(w_x-\mu_i) \mathcal{N}(w_e; \boldsymbol 0, \Sigma).
	\end{align*}
	Both integral solutions explain that the marginalization leads to the distributions of $\mathcal{Q}_w$ and $\mathcal{Q}_w^x \otimes \mathcal{Q}_w^e$, proving our claim.
	
	With the preliminaries established, we claim system $\hat \M$ is simulated by system $\M$. To prove this claim, we consider a relation between $x$, $x_i$, $z$, $e$ and $\hat x_i$ given by $\mathcal{R}_2:=\{(z,e,\hat x_i, x, x_i) \mid x=z+e, x_i=\hat x_i\}$; establish an interface function $\nu(e,v)=v+Ke$; and derive a stochastic kernel $\mathbb{W}_2(\,\cdot\,{\mid} x,x_i,z,e,\hat x_i,v)$ from the lifting $\mathbb{W}_1$. Since $x,z,e,\hat x_i$ of relation $\mathcal{R}_2$ are each directly related to an element in relation $\mathcal{R}_1$, and $x_i$ is directly determined from the interface function, any result in regards to Def. \ref{Def:lifting} on $\mathbb{W}_1$ can be extended towards $\mathbb{W}_2$. Consequently, the satisfaction of the second requirement of Def. \ref{Def:SSR} follows from the fact that the stochastic kernel $\mathbb{W}_2$ can be derived from the lifting $\mathbb{W}_1$ and shares the same results as $\mathbb{W}_1$ in regards to Def. \ref{Def:lifting}, only now extended towards the transition kernels. Since the first requirement is trivially satisfied, our claim is proven.
\end{proof}

With the SSR between $\M$ and $\hat{\M}$ established, we consider the following proposition from \cite[Thm. 2]{Haesaert2017verification}.

\begin{proposition} \label{Prop:SSR}
	If $\M_1$ is simulated by $\M_2$ then for all control strategies $\boldsymbol f_1$ there exists a control strategy $\boldsymbol f_2$ such that, for all measurable events $\mathcal{A} \in \mathcal{B}(\mathbb{Y}^{N+1})$, $$\mathbb{P}_{\boldsymbol f_1 \times \M_1}(\{h_1(x_1)\}_{[0,N]} \in \mathcal{A}) = \mathbb{P}_{\boldsymbol f_2 \times \M_2}(\{h_2(x_2)\}_{[0,N]} \in \mathcal{A}),$$
	where $\{h_j(x_j)\}_{[0,N]} := \{h_j(x_j(0)), \cdots, h_j(x_j(N))\}$.
\end{proposition}


\subsection{Control Refinement Operator}

\begin{figure}[htp]
	\centering
	\includegraphics[width=\columnwidth]{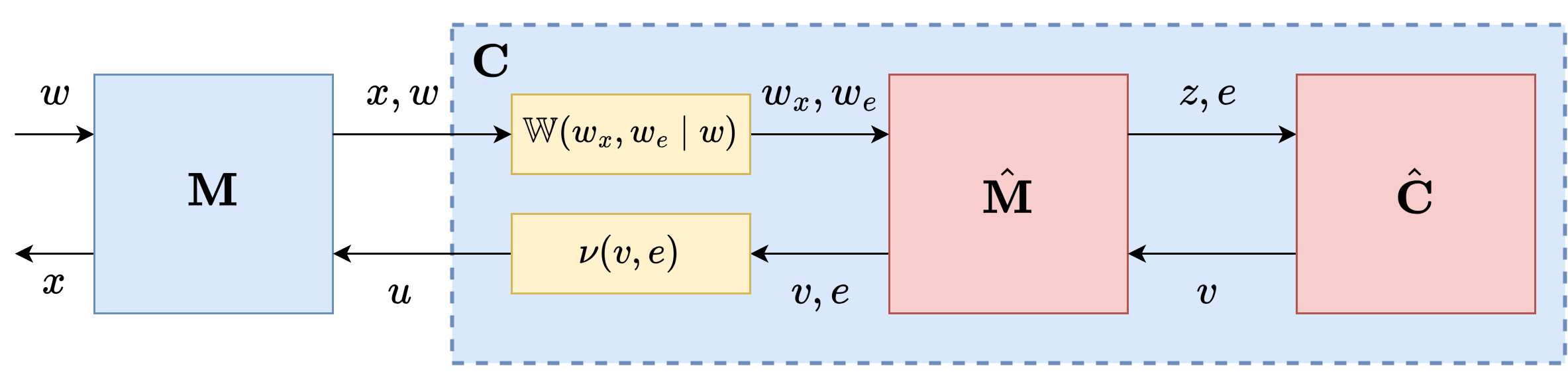}
	\caption{An illustration of the control refinement operator. The original system is denoted by $\M$, the abstract system by $\hat \M$, the abstract control strategy by $\hat \C$ and the original control strategy by $\C$.}
	\label{Fig:ConRef}
\end{figure}

Since Prop. \ref{Prop:SSR} guarantees there exists a CR-operator that preserves chance constraint satisfaction, we reformulate the SMPC problem \eqref{SMPC} onto the abstract system:
\begin{subequations} \label{SMPC_Abs}
	\begin{align}
		&\textstyle \min _{\boldsymbol{\hat{f}_k}} \quad J(\boldsymbol z_k, \boldsymbol v_k), \quad \text{s.t.} \\
		&\hspace{-0.3em}z(k+i+1)  = Az(k+i)+Bv(k+i) + w_x(k+i),\\
		&\hspace{-0.3em}e(k+i+1) = A_Ke(k+i)+w_e(k+i),\\
		&\hspace{-0.3em}z(k+i+1)+e(k+i+1) \text{ satisfy } \eqref{ChanCons_x}, \\
		&\hspace{-0.3em}v(k+i)+Ke(k+i) \text{ satisfy } \eqref{ChanCons_u}, \ \forall i \in \mathbb{N}.
	\end{align}
\end{subequations}
Next, we design a CR-operator $D:\boldsymbol{\hat{f}_k} \to \boldsymbol f_k$ by utilizing the lifting and interface function. Based on Thm. \ref{Thm:SSR}, for any realization $w \sim \mathcal{Q}_w$, there exists a conditional probability measure given by
\begin{align} \label{Eq:CondProb}
	&\textstyle \mathbb{W}(w_x,w_e \mid w) := \mathbb{W}_1(w_x,w_e \mid w; \boldsymbol \mu, \boldsymbol \pi, \Sigma) =\nonumber\\
	&\textstyle\sum_i\left[ \delta(w_x-\mu_i)\frac{\pi_i \mathcal{N}(w; \mu_i, \Sigma)}{\GaussMix(w;\boldsymbol \mu, \boldsymbol \pi, \Sigma)}\right] \delta(w_e-w+w_x),
\end{align}
such that any realization of \eqref{Eq:CondProb} ensures both the original and abstract system have the same output if $u(k)=v(k)+Ke(k)$. Accordingly, the interface function and conditional probability measure \eqref{Eq:CondProb} establish a CR-operator that maps a control strategy synthesized on the abstract system into a control strategy on the original system. Fig. \ref{Fig:ConRef} illustrates the control refinement operator.

\section{Control Synthesis for the Abstract System} \label{Sec:Main_Con}

This section presents an approach to solving the SMPC problem \eqref{SMPC_Abs}, building on \cite{Hewing2018,Engelaar2023stochastic}. We define probabilistic reachable sets to tighten the chance constraints \eqref{ChanCons}, reformulating them as deterministic constraints on the nominal dynamics while eliminating error dynamics. To account for residual stochasticity, we introduce a branch model predictive control framework, similar to \cite{chen2022interactive}, where each branch represents a realization of the residual uncertainty.


\subsection{Bounding the Error Dynamics}

As is common, we bound the effect of the error dynamics  $e(k)$ in \eqref{AbsSys}. Afterwards, we tighten the constraints on the nominal dynamics $z(k)$. To accomplish both, we define probabilistic reachable sets over the error dynamics \eqref{Eq:Err}.
\begin{definition}[Probabilistic Reachable Sets \cite{Hewing2018}] \label{Def:PRS}
	A probabilistic reachable set (PRS) of probability level $p \in [0,1]$ for error dynamics \eqref{Eq:Err}, denoted by $\mathcal{R}^p$, is a set that satisfies
	\begin{equation}\label{Eq:PRT}
		e(0) = \boldsymbol 0 \Rightarrow \Pr(e(k) \in \mathcal{R}^p) \geq p, \ \forall k \geq 0.
	\end{equation}
\end{definition}
\smallskip

The above definition states that a PRS of probability level $p$ will contain at any time instance $k$ the accumulating error $e(k)$ with at least probability $p$, if the initial error satisfies $e(0)=\boldsymbol 0$. Multiple approaches exist to obtaining an explicit form for any given PRS; see \cite{Hewing2018,Hewing2020}. As in \cite[Sec. 3.2]{Hewing2020}, we consider the ellipsoidal explicit representation, given by
\begin{equation}\label{EllipNot}
	\mathcal{R}^p= \{x \in \mathbb{R}^{n} \mid x^T \Sigma_{\infty}^{-1}x \leq \tilde p\},
\end{equation}
where $\Sigma_{\infty}$ solves the Lyapunov equation $A_K\Sigma_{\infty}A_K^T+\text{var}(\mathcal{Q}_w)=\Sigma_{\infty}$ and $\tilde p=\chi^2_{n}(p)$ where $\chi^2_{n}(\cdot)$ is the inverse cumulative distribution function of the chi-squared distribution with $n$ degrees of freedom. Since the variance of $\mathcal{Q}_w$ is strictly positive definite and $A_K$ is stable, Lyapunov theory states a strictly positive definite $\Sigma_{\infty}$ exists \cite{Datta2004numerical}.


\subsection{Tightening the Chance Constraints}

Based on the probabilistic reachable sets $\mathcal R^p$, we can decompose and tighten the chance constraints \eqref{ChanCons_x} on $z(k)+e(k)$ into a constraint for the error dynamics $e(k)$ and a constraint for the nominal state $z(k)$, since
\begin{equation*}
    \Pr( z(k)+e(k)\in \mathcal{X})\geq \Pr(z(k) \in \mathcal{X}\ominus \mathcal{R}_x\wedge e(k) \in \mathcal{R}_x)\geq p_x,
\end{equation*}
if $\mathcal{R}_x := \mathcal R^{p_x}$ (c.f. \eqref{EllipNot}) and $\Pr(z(k) \in \mathcal{X}\ominus \mathcal{R}_x) = 1$. The result is the deterministic constraint $z(k) \in \mathcal{X}\ominus \mathcal{R}_x$ on the nominal state. 
Similarly, we can decompose and tighten the chance constraint \eqref{ChanCons_u} on $v(k)+Ke(k)$ into a deterministic constraint on $v(k)$. All in all, the chance constraints can be replaced by the deterministic constraints
\begin{subequations} \label{DynDetCon}
	\begin{align}
		z(k) &\in \mathcal{Z}:=\mathcal{X}\ominus \mathcal{R}_x, \ \mathcal{R}_x = \mathcal R^{p_x}, \label{DynDetCon_Stat}\\
		v(k) &\in \mathcal{V}:=\mathcal{U}\ominus K\mathcal{R}_u, \ \mathcal{R}_u = \mathcal R^{p_u}.
	\end{align}
\end{subequations}
Due to the chosen ellipsoidal representation, the above tightening introduces conservatism.

In the sequel, we will compute with \eqref{DynDetCon_Stat} as though $\mathbb{P}(z(k) \in \mathcal{Z})=p_z=1$. However, we would like to remark that probability $p_z$ can be relaxed if $\mathcal{R}_x=\mathcal R^{p_e}$ with $p_ep_z\geq p_x$. The same holds for the nominal input $v(k)$.


\subsection{BMPC Design on the Nominal Dynamics}

With tightening ensuring that the error dynamics \eqref{Eq:Err} and chance constraints \eqref{ChanCons} are transformed into deterministic constraints \eqref{DynDetCon}, the remaining challenge is the discrete distribution in the nominal dynamics \eqref{Eq:Nom}. To address this, we design a control strategy for each realization of the nominal disturbance. This approach resembles contingency planning \cite{ren2024recursively}, but unlike contingency planning, where a single strategy applies to all realizations, we design a separate strategy for each realization. To achieve this, we adopt a branch model predictive control framework akin to \cite{chen2022interactive}.

\begin{figure}[htp]
    \centering
    \includegraphics[width=\columnwidth]{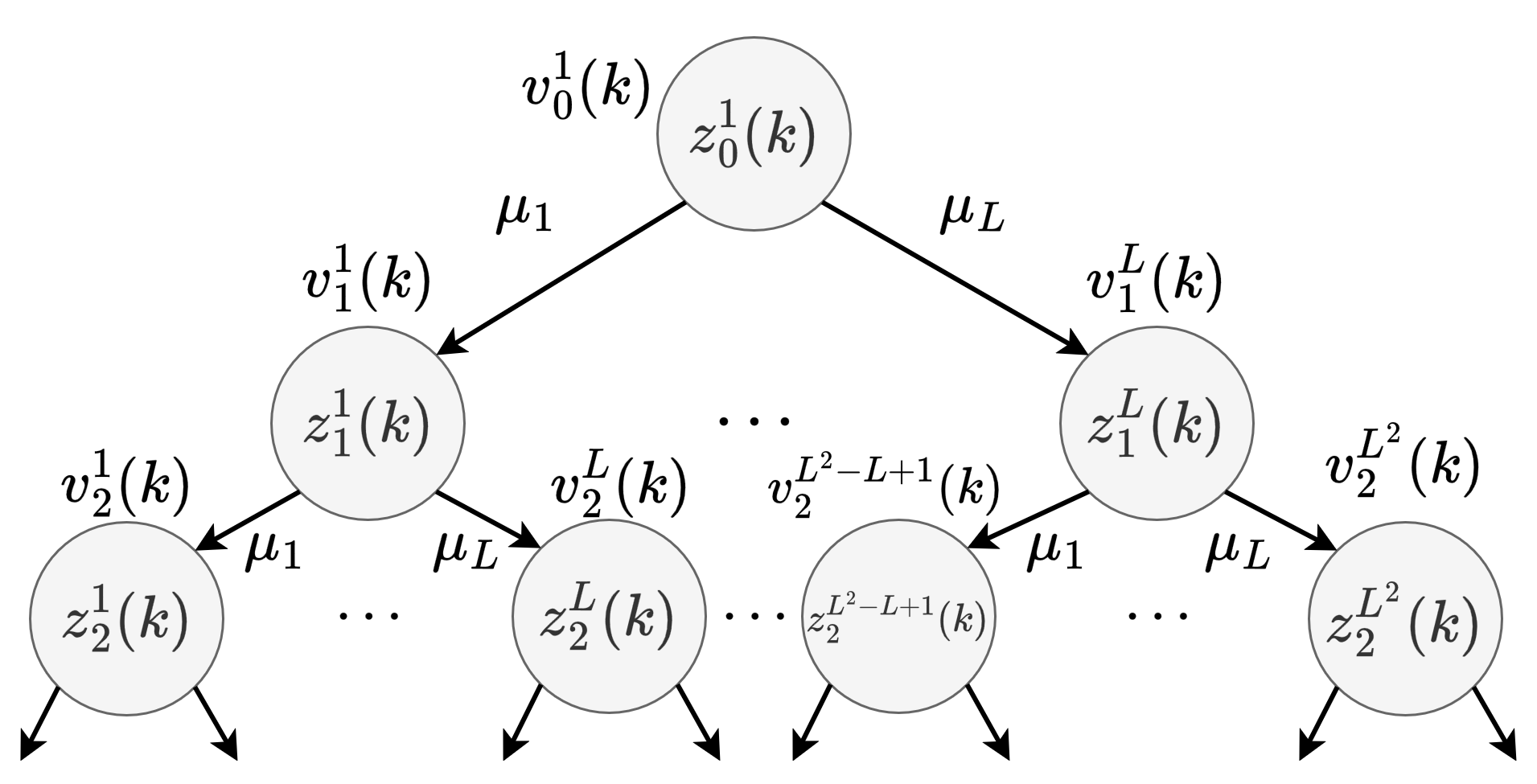}
    \caption{Each branch is a potential control strategy, corresponding to a realization of $\mathcal{Q}_w^x \otimes \cdots \otimes \mathcal{Q}_w^x$.}
    \label{Fig:Tree}
\end{figure}

Under the assumption of a finite time horizon $N$ and a finite set of nominal disturbances given by $\mathcal{W}:=\{\mu_1, \cdots, \mu_L\}$, we design a finite tree of $L^N$ branches, where each branch represents a potential control strategy. Each strategy corresponds to a unique sequence $\{\mu_{i_1}, \cdots, \mu_{i_N}\}$, $i_j \in \{1, \cdots, L\}$, capturing all nominal disturbance behaviour of length $N$. Fig. \ref{Fig:Tree} gives a visualization of the tree.

We denote for any vector or scalar of sequences $\boldsymbol s(k)$, the corresponding predicted value of $s(k+i)$ by $s_i(k)$. Regarding the branch superscript, we follow the pattern illustrated in Fig. \ref{Fig:Tree}. To synthesize a control strategy for each disturbance realization, we define the following branch MPC problem.
\begin{subequations} \label{Eq:MPC}
    \begin{align}
        &\textstyle \min _{\bar z(k), \bar v(k), \xi(k)} \quad J(\bar z(k), \bar v(k)) + \epsilon \xi(k)^2 \\[.8em]
        & \text{s.t. } z^{(j-1)L+d}_{i+1}(k)=A z^j_{i}(k)+Bv^j_{i}(k)+\mu_d,\label{Eq:MPCDyn}\\ 
        & z^j_{i}(k)\in \mathcal{Z}, \ v^j_{i}(k)\in \mathcal{V}, \ z^l_{N}(k)\in \mathcal{Z}_f, \label{Eq:MPCSet} \\
        & z^1_0(k) = (1-\xi(k))x(k)+\xi(k) z^{c}_1(k-1), \label{Eq:MPCIni}\\
        & \xi(k) \in \{0,1\}, \ \forall i \in \{0,\ldots, N-1\}, \label{Eq:MPCXi} \\
        & \forall d \in \{1,\ldots,L\}, \forall j \in \{1,\ldots, L^i\}, \forall l \in \{1, \ldots, L^N\}, \nonumber 
    \end{align}
\end{subequations}
where $\bar z(k):=\{z_0^1(k), z_1^1(k), \cdots, z_N^{L^N}(k)\}$ and $\bar v(k):=\{v_0^1(k), v_1^1(k), \cdots, v_{N-1}^{L^{N-1}}(k)\}$ are sequences of nominal states and inputs, as observed in Fig. \ref{Fig:Tree}; $z_1^c(k-1)$ is the previously computed nominal state and $z_1^c(-1)=x(0)$; $J$ is a linear or quadratic function; and $\mathcal{Z}_F$ is the terminal set. 

The terminal set, with respect to input $v(k)=Kz(k)$, must satisfy $(A+BK)\mathcal{Z}_F \oplus \mathcal{W} \subseteq \mathcal{Z}_F \subseteq \mathcal{Z}$ and $K\mathcal{Z}_F \subseteq \mathcal{V}$ to ensure recursive feasibility. Should set $\mathcal{Z}$ and $\mathcal{V}$ be (under-approximated by) polyhedral sets, a terminal set $\mathcal{Z}_F$ can be obtained from \cite[Thm. 3.1]{Kouvaritakis2016}. A more general algorithm to compute the terminal set is given by \cite[Alg. 10.4]{borrelli2017predictive}. The choice of initial state $z_0^1(k)$ comes from the idea put forward in \cite{Hewing2018}. Therein, both the real measurement $x(k)$ and the previously computed nominal state $z_1^c(k-1)$ can become the current initial state, again to ensure recursive feasibility. 

\begin{remark}
     The number of optimization variables in MPC problem \eqref{Eq:MPC} grows exponentially with the number of mixture components and the time horizon. In future work, we opt to reduce the complexity. To accomplish this, we consider the work by \cite{Goulart2006optimization}, wherein the growth can be reduced to (potentially) quadratic under the assumption of polyhedral nominal set constraints. However, only a single control strategy will be synthesized for all branches, leading to a conservative control strategy. An alternative second approach considers the mixed logic dynamical framework described in \cite{Bemporad2000predictive}. This framework, too, can potentially reduce the complexity at the cost of solving a mixed integer quadratic programming problem. A third approach considers scenario MPC, which fixes the disturbance after a specified time \cite{Lucia2013multi}, reducing the number of branches. In addition to the previously mentioned approaches, we might also consider removing branches, thereby reducing the probability of $\Pr(z(k)+\mathcal{R}_x \subseteq \mathcal{X})$ and $\Pr(v(k)+K\mathcal{R}_x \subseteq \mathcal{U})$. To ensure the satisfaction of \eqref{ChanCons}, the trade-off implies tighter constraints \eqref{DynDetCon}.
\end{remark}


\section{Implementation \& Closed-loop Guarantees} \label{Sec:Analysis}

In this section, we determine recursive feasibility of \eqref{Eq:MPC}, design a control strategy implementation algorithm for the original system and establish that the closed-loop dynamics of the original system will satisfy the chance constraints \eqref{ChanCons}.

\begin{theorem}[Recursive Feasibility]\label{Thm:RecFeas}
The BMPC problem \eqref{Eq:MPC} has a solution for all time $k \in \mathbb{N}$, assuming that a solution exists at time $k=0$.
\end{theorem} 

\begin{proof}
Considering proof by induction, we assume 
\begin{align*}
    \bar z(k) &= \{z_0^1(k), z_1^1(k), \cdots z_N^{L^N}(k)\},\\
    \bar v(k) &= \{v_0^1(k), v_1^1(k), \cdots v_{N-1}^{L^{N-1}}(k)\}, 
\end{align*}
is a solution of \eqref{Eq:MPC} at time $k$, and, without loss of generality, that $z_1^c(k)=z_1^j(k)$ for some $j \in \{1, \cdots, L\}$. We take $\xi(k+1)=1$, and $\bar z(k+1)$ and $\bar v(k+1)$ to be defined by the following set of relations, given by
\begin{subequations}
    \begin{align}
        z_{0}^1(k+1) &= z_1^j(k),\\
        z_{i}^{(l-1)L+d}(k+1) &= z_{i+1}^{(j-1)L^i+(l-1)L+d}(k), \label{Eq:PrefSta} \\
        z_{N}^{(l-1)L+d}(k+1) &= A_Kz_{N-1}^{l}(k+1)+\mu_d, \ i=N,\\
        v_{0}^1(k+1) &= v_1^j(k), \\
        v_{i}^{(l-1)L+d}(k+1) &= v_{i+1}^{(j-1)L^i+(l-1)L+d}(k), \label{Eq:PrefInp} \\
        v_{N-1}^{l}(k+1) &= Kz_{N-1}^{l}(k+1), \ i=N,
    \end{align}
\end{subequations}
for $l \in \{1, \cdots, L^{i-1}\}$, $i \in \{1, \cdots, N-1\}$ and $d \in \{1, \cdots, L\}$. When considering the tree at time $k$, prefixes \eqref{Eq:PrefSta} and \eqref{Eq:PrefInp} represent, respectively, the nominal state and input corresponding to the sub-tree with origin $z_1^j(k)$.

We note that $\bar z(k+1)$, $\bar v(k+1)$ and $\xi(k+1)$ satisfy conditions \eqref{Eq:MPCDyn}, \eqref{Eq:MPCIni} and \eqref{Eq:MPCXi} trivially at time $k+1$. Similarly, condition \eqref{Eq:MPCSet} is trivially satisfied for all $i\in \{0, \cdots, N-2\}$. Regarding $i=N-1$ and the terminal set constraint, satisfaction is a result of the conditions $(A+BK)\mathcal{Z}_F \oplus \mathcal{W} \subseteq \mathcal{Z}_F \subseteq \mathcal{Z}$ and $K\mathcal{Z}_F \subseteq \mathcal{V}$, and the observation that $z_{N-1}^l(k+1) \in \mathcal{Z}_F$ for all $l \in \{1, \cdots, L^{N-1}\}$. Thus, a solution for MPC problem \eqref{Eq:MPC} exists at time $k+1$ if a solution exists at time $k$. This finishes the proof.
\end{proof}

Algorithm \ref{Alg:Implement} outlines the control implementation for the original system \eqref{Sys}, demonstrating how recursive control is achieved by utilizing the control refinement operator.

\begin{algorithm}[ht]
    \caption{Implementation Algorithm} \label{Alg:Implement}
	\begin{algorithmic}[1]
		\State Given: Dynamics \eqref{Sys}, $x(0)$, $\mathcal{X}$, $\mathcal{U}$, $p_x$, $p_u$, $J(\cdot)$, and $\epsilon$
        \State Construct $\mathcal{R}_x$ and $\mathcal{R}_u$ and compute $\mathcal{Z}$, $\mathcal{V}$ and $\mathcal{Z}_F$
        \For{each time instance $k \in \mathbb{N}$}
        \State Solve \eqref{Eq:MPC} to obtain $z_0^1(k)$ and $v_0^1(k)$
        \State Compute $e(k)$ from \eqref{Eq:Err} and $u(k)$ from \eqref{RelInput}
        \State Obtain $x(k+1)$ and compute $w(k)$ from \eqref{Sys}
        \State Draw $\mu_j=w_x$ from \eqref{Eq:CondProb} and set $z_1^c(k)=z_1^j(k)$
        \EndFor
	\end{algorithmic}
\end{algorithm}

The following proposition establishes satisfaction of the chance constraints \eqref{ChanCons} for the closed-loop abstract system.
\begin{proposition}[PRS for closed-loop error] \label{Prop:Error}
    Let $\mathbf{\mathcal{R}}$ be a convex symmetric set. For system \eqref{AbsSys} under the control input $v(k)=v_0^1(k)$ resulting from Alg. \ref{Alg:Implement}, we have that
    \begin{equation}
        \mathbb{P}(e_0(k) \in \mathcal{R})\geq \mathbb{P}(e_k(0) \in \mathcal{R}), \quad \forall k \geq 0.
    \end{equation}
\end{proposition}
\smallskip
The proof of Prop. \ref{Prop:Error} can be found in \cite[Thm. 3]{Hewing2018}. Though \cite{Hewing2018} does not consider a BMPC framework, the result of \cite[Thm. 3]{Hewing2018} remains valid within this framework as its proof remains accurate regardless of which branch is implemented at each time instance. This is because the proof regards only the error dynamics, which remain consistent with \cite{Hewing2018} within this paper. 

Prop. \ref{Prop:Error} implies that $\mathbb{P}(e(k) \in \mathcal{R}_x) \geq p_x$ and $\mathbb{P}(Ke(k) \in \mathcal{R}_u) \geq p_u$, $\forall k \in \mathbb{N}$. Hence, the abstract system under the control input $v(k)=v_0^1(k)$ resulting from Alg. \ref{Alg:Implement} satisfies chance constraints \eqref{ChanCons}. Consequently, we have the following corollary, the proof of which follows from Prop. \ref{Prop:SSR}.
\begin{corollary}
    System \eqref{Sys} under the control law resulting from Alg. \ref{Alg:Implement} will satisfy the chance constraints \eqref{ChanCons}.
\end{corollary}


\section{Case Study} \label{Sec:Case}

To illustrate that our control synthesis algorithm allows for the satisfaction of the chance constraints, we consider a vehicle control case study in which the vehicle must maintain its position on an ill-maintained road. The vehicle's position on the road is defined by a one-dimensional disturbed integrator dynamical system given by $x(k+1)=x(k)+u(k)+w(k)$ where $w(k) \sim \GaussMix(\boldsymbol \mu,\boldsymbol \pi, \Sigma)$ represents the vehicle's deviation as a result of the road's condition. We assume that $\boldsymbol{\mu}=\{-1.5,0,1.5\}$, $\boldsymbol{\pi}=\{0.2,0.3,0.5\}$ and $\Sigma=0.25$. The road is split into an inner and outer part, each of which the vehicle must remain within with a certain probability. The inner road is defined by a convex set $\mathcal{X}_{Inn}=\{x \mid -2 \leq x \leq 2\}$ and the target lower bound $p_{Inn}=0.6$. The outer road is defined by a convex set $\mathcal{X}_{Out}=\{x \mid -3 \leq x \leq 3\}$ and the target lower bound $p_{Out}=0.99$. Additionally, the velocity of the vehicle is also restricted. This restriction is represented on the input space via the convex set $\mathcal{U}=\{u \mid -2 \leq u \leq 2\}$ and the target lower bound $p_u=0.65$.

We consider a total horizon of 10 steps and an MPC horizon of 5 steps. After simulating 1000 control strategies for differing realizations of the Gaussian mixture disturbance, we obtain the following graphs regarding the vehicle's behaviour and the number of constraint violations. As can be extrapolated from the graphs, the vehicle satisfies the inner road, outer road and velocity constraints with at least probability equal to the target lower bounds. Moreover, it can be inferred that the control strategies are conservative as all chance constraints are easily met. Based on the data, the inner road chance constraint would be satisfied with probability $0.86$, the outer road with probability $0.99$ and the velocity with probability $0.89$.

\begin{figure}[htp]
	\centering
	\includegraphics[width=\columnwidth]{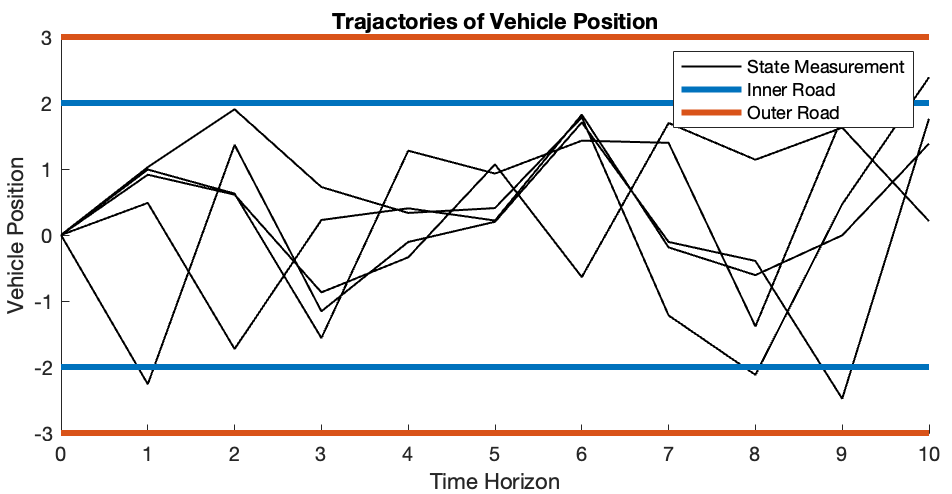}
	\caption{A line graph showing a few trajectories of the vehicle. Each trajectory corresponds to different realizations of the Gaussian mixture.}
	\label{Fig:resultState}
\end{figure}

\begin{figure}[htp]
	\centering
	\includegraphics[width=\columnwidth]{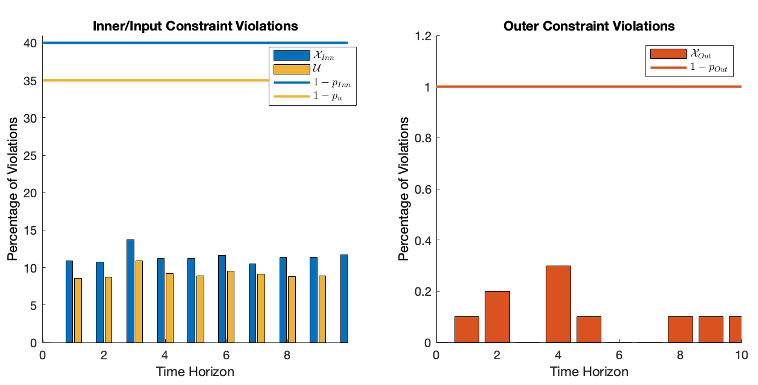}
	\caption{A bar graph of the percentage of constraint violations after 1000 simulations. The lines indicate the upper bounds of the road and velocity violations in accordance with the target lower bounds.}
	\label{Fig:resultViolation}
\end{figure}

We utilized a MacBook Pro (14-inch, 2021) with an Apple M1 Pro processor (8-core CPU: 6 performance cores and 2 efficiency cores), a 14-core integrated Apple GPU and 16 RAM Memory. We ran our code in the Matlab environment and utilized a Gurobi solver. After offline preprocessing, on average, a single online simulation took $380\sim410$ ms. Each simulation contained 16767 1-D linear constraints and 488 optimization variables.


\section{Conclusion}

In conclusion, we addressed the control synthesis problem for linear systems subject to additive Gaussian mixture disturbances to satisfy chance constraints while preserving key properties such as recursive feasibility and closed-loop guarantees. This was achieved by formulating an abstract system, redefining the SMPC problem onto the abstract system, and reformulating the redefined SMPC problem into a BMPC framework. Our main contribution is the extension of the SMPC methods towards Gaussian mixtures. While our method does contain conservativeness, we want to emphasise that this is partially inhered from the BMPC reformulation of the redefined SMPC problem. Future research may enhance the framework by considering or developing more computationally efficient methods to solve the MPC problem.


\bibliographystyle{IEEEtran}
\bibliography{References}

\end{document}